\documentclass[aps,prd,twocolumn,superscriptaddress]{revtex4}
\usepackage{epsfig,epsf}
\usepackage{amsmath}
\usepackage{amsthm}
\usepackage{amsfonts}
\usepackage{amssymb}
\usepackage{dsfont}
\usepackage{multirow}
\usepackage{appendix}
\usepackage{slashed}
\usepackage[active]{srcltx}
\usepackage{psfrag}

\begin{document}

\title{Stable scalar tetraquark $T_{bb;\bar{u}\bar{d}}^{-}$ }
\author{S.~S.~Agaev}
\affiliation{Institute for Physical Problems, Baku State University, Az--1148 Baku,
Azerbaijan}
\author{K.~Azizi}
\affiliation{Department of Physics, University of Tehran, North Karegar Avenue, Tehran
14395-547, Iran}
\affiliation{Department of Physics, Do\v{g}u\c{s} University, Acibadem-Kadik\"{o}y, 34722
Istanbul, Turkey}
\affiliation{School of Particles and Accelerators, Institute for Research in Fundamental
Sciences (IPM) P.O. Box 19395-5531, Tehran, Iran}
\author{B.~Barsbay}
\affiliation{Department of Physics, Do\v{g}u\c{s} University, Acibadem-Kadik\"{o}y, 34722
Istanbul, Turkey}
\affiliation{Department of Physics, Kocaeli University, 41380 Izmit, Turkey}
\author{H.~Sundu}
\affiliation{Department of Physics, Kocaeli University, 41380 Izmit, Turkey}

\begin{abstract}
The mass and coupling of the scalar tetraquark $T_{bb;\overline{u}\overline{d%
}}^{-}$ (hereafter $T_{b:\overline{d}}^{-} $) are calculated in the context
of the QCD two-point sum rule method. In computations we take into account
effects of various quark, gluon and mixed condensates up to dimension ten.
The result obtained for the mass of this state $m=(10135\pm 240)~\mathrm{MeV}
$ demonstrates that it is stable against the strong and electromagnetic
decays. We also explore the dominant semileptonic $T_{b:\overline{d}}^{-}
\to \widetilde{Z}_{bc;\bar{u}\bar{d}}^{0}l\overline{\nu }_{l}$ and
nonleptonic decays $T_{b:\overline{d}}^{-} \to \widetilde{Z}_{bc;\bar{u}\bar{%
d}}^{0}M$, where $\widetilde{Z}_{bc;\bar{u}\bar{d}}^{0}$ is the scalar
tetraquark composed of color-sextet diquark and antidiquark, and $M$ is one
of the final-state pseudoscalar mesons $\pi^{-}, K^{-}, D^{-}$ and $D_s^{-}$%
, respectively. The partial widths of these processes are calculated in
terms of the weak form factors $G_{1(2)}(q^2)$, which are determined from
the QCD three-point sum rules. Predictions for the mass, full width $\Gamma
_{\mathrm{full}} =(10.88\pm 1.88)\times 10^{-10}~\mathrm{MeV}$, and mean
lifetime $\tau=0.61_{-0.09}^{+0.13}~\mathrm{ps}$ of the $T_{b:\overline{d}%
}^{-}$ obtained in the present work can be used in theoretical and
experimental studies of this exotic state.
\end{abstract}

\maketitle

%%%%%%%%%%%%%%%%%%%%%%%%%%%%%%%%%%%%%%%%%%%%%%%%%%%%%%%%%%%%%%

\section{Introduction}

\label{sec:Int}
%%%%%%%%%%%%%%%%%%%%%%%%%%%%%%%%%%%%%%%%%%%%%%%%%%%%%%%%%%%%%%
Four-quark states composed of heavy diquarks $QQ^{\prime }$ and light $%
\overline{q}\overline{q}^{\prime }$ antidiquarks are real candidates to
stable exotic mesons. During last few years interest to these tetraquarks is
renewed, although main qualitative results concerning a stability of the
compounds $QQ^{\prime }\overline{q}\overline{q}^{\prime }$ against strong
decays were obtained many years ago \cite%
{Ader:1981db,Lipkin:1986dw,Zouzou:1986qh,Carlson:1987hh}. Thus, it was shown
that such four-quark mesons would be stable if the ratio $m_{Q}/m_{q}$ is
sufficiently large. A prominent particle from this series is the
axial-vector state $T_{bb;\overline{u}\overline{d}}^{-}$ (in what follows $%
T_{bb}^{-}$): Studies conducted in the framework of different models
confirmed, that its mass is below the $B\overline{B}^{\ast }$ threshold, and
$T_{bb}^{-}$ is the particle stable against strong decays \cite%
{Carlson:1987hh,Navarra:2007yw}.

Discovery of double-charmed $\Xi _{cc}^{++}$ baryon stimulated
investigations of heavy tetraquarks because parameters of this baryon were
used in phenomenological models to estimate the mass of $T_{bb}^{-}$ \cite%
{Karliner:2017qjm,Eichten:2017ffp}. The prediction for the mass of $%
T_{bb}^{-}$ obtained in Ref.\ \cite{Karliner:2017qjm} equals to $m=(10389\pm
12)~\mathrm{MeV}$ being $215~\mathrm{MeV}$ below $B^{-}\overline{B}^{\ast 0}$
and $170~\mathrm{MeV}$ below $B^{-}\overline{B}^{0}\gamma $ thresholds,
respectively. This means that the tetraquark $T_{bb}^{-}$ is stable against
the strong and radiative decays and should dissociate to conventional mesons
only weakly. The similar conclusion on strong-interaction stable nature of $%
T_{bb}^{-}$ was made in Ref.\ \cite{Eichten:2017ffp} on the basis of the
heavy-quark symmetry analysis. Its mass was found equal to $m=10482~\mathrm{MeV}$
which is $121~\mathrm{MeV}$ below the open-bottom threshold.

In the context of the QCD sum rule method the axial-vector particle $%
T_{bb}^{-}$ was recently studied in our paper \cite{Agaev:2018khe}. In
accordance with Ref.\ \cite{Agaev:2018khe} the mass of $T_{bb}^{-}$ amounts
to $m=(10035~\pm 260)~\mathrm{MeV}$ that confirms once more its stability
against the strong and radiative decays. We also explored the semileptonic
decays $T_{bb}^{-}$ $\rightarrow Z_{bc}^{0}l\overline{\nu }_{l}$ and
calculated partial widths of these processes. In these decays, we treated
the final-state tetraquark $Z_{bc}^{0}=[bc][\overline{u}\overline{d}]$ as a
scalar particle built of color-triplet diquarks $[\overline{\mathbf{3}}%
_{c}]_{bc}\otimes \lbrack \mathbf{3}_{c}]_{\overline{u}\overline{d}}$. The
predictions for the full width $\Gamma =(7.17\pm 1.23)\times 10^{-8}~\mathrm{%
MeV}$ and mean lifetime $\tau =9.18_{-1.34}^{+1.90}~\mathrm{fs}$ of the
axial-vector tetraquark $T_{bb}^{-}$ are useful for experimental
investigation of a family of double-heavy exotic mesons. The parameters of
the $T_{bb}^{-}$ and its weak decays were considered in Ref.\ \cite%
{Hernandez:2019eox} as well.

It is worth noting that apart from $T_{bb}^{-}$, some of tetraquarks
containing heavy $bb$ and $bc$ diquarks and light antidiquarks may be stable
against strong (radiative) decays, and transform to ordinary mesons via weak
interactions. Exotic mesons with various quantum numbers composed of \ heavy
$bb$ or $bc$ diquarks were also objects of intensive studies. Thus, the
parameters of the four-quark compounds $QQ\overline{q}\overline{q}$ with the
spin-parities $J^{P}=0^{-},\ 0^{+},\ 1^{-}$ and $1^{+}$ were evaluated in
the context of the QCD sum rule method in Ref.\ \cite{Du:2012wp}. We
explored the heavy exotic scalar meson $T_{bb;\overline{u}\overline{s}}^{-}$%
, and calculated its mass, width and mean lifetime \cite{Agaev:2019lwh}. The
charged tetraquarks $Z_{bc;\overline{u}\overline{u}}^{-}$ and $Z_{bc;%
\overline{d}\overline{d}}^{+}$ were investigated in Ref.\ \cite{Chen:2013aba}%
, and the prediction $m=(7.14\pm 0.10)~\mathrm{GeV}$ was made for their
masses. The scalar and axial-vector states $bc\overline{u}\overline{d}$
were\ in the focus of theoretical studies as well. Indeed, calculations
carried out in Ref.\ \cite{Karliner:2017qjm} proved that $Z_{bc}^{0}$ is
below the threshold for $S$-wave decays to ordinary mesons $B^{-}D^{+}$ and $%
\overline{B^{0}}D^{0}$. To similar conclusions led analysis of the
ground-state $QQ^{\prime }\overline{u}\overline{d}$ tetraquarks' masses
performed using the Bethe-Salpeter method \cite{Feng:2013kea}: the mass of $%
Z_{bc}^{0}$ found there equals to $6.93~\mathrm{GeV}$ and is lower than the
relevant strong threshold. The lattice simulations showed the
strong-interaction stability of the $I(J^{P})=0(1^{+})$ exotic meson $Z_{ud;%
\overline{c}\overline{b}}^{0}$ with the mass in the range $15$ to $61$ $%
\mathrm{MeV}$ below $\overline{D}B^{\ast }$ threshold \cite{Francis:2018jyb}%
. Another confirmation of the stable nature of the tetraquarks $bc\overline{u%
}\overline{d}$ came from Ref.\ \cite{Caramees:2018oue}. In this article it
was shown that both the $J^{P}=0^{+}$ and $1^{+}$ isoscalar tetraquarks $bc%
\overline{u}\overline{d}$ are stable against the strong decays. The
isoscalar tetraquark with $J^{P}=0^{+}$ is also electromagnetic-interaction
stable particle, whereas $J^{P}=1^{+}$ may transform to the final state $%
\overline{B}D\gamma $ through the electromagnetic interaction. On the
contrary, the masses of the scalar and axial-vector $bc\overline{u}\overline{%
d}$ states were estimated respectively around $7229~\mathrm{MeV}$ and $7272~%
\mathrm{MeV}$, which mean that they can decay to ordinary mesons $B^{-}D^{+}/%
\overline{B^{0}}D^{0}$ and $B^{\ast }D$ \cite{Eichten:2017ffp}.

We determined spectroscopic parameters of the scalar exotic meson $Z_{bc}^{0}
$ in Ref.\ \cite{Agaev:2018khe}. To this end, we used the QCD sum rule
method and found $m_{Z}=(6660\pm 150)~\mathrm{MeV}$. This result is
considerably below thresholds for strong and radiative decays of $Z_{bc}^{0}$
to conventional heavy mesons $B^{-}D^{+}$ and $\overline{B^{0}}D^{0}$, and
to final states $\overline{B^{0}}D_{1}^{0}\gamma $ and $B^{\ast }D_{0}^{\ast
}\gamma $, respectively. In other words, the scalar tetraquark $Z_{bc}^{0}$
is the strong- and electromagnetic-interaction stable particle, therefore
its weak decay channels are of special interest \cite{Sundu:2019feu}. The
axial-vector exotic meson $[bc][\overline{u}\overline{d}]$ was investigated in Ref.\ \cite{Agaev:2019kkz}, in which its spectroscopic parameters and possible strong and weak decays were considered.

Four quark mesons $T_{cc}$ made of diquarks $cc$ and light antidiquarks,
their properties and decays were studied in numerous publications \cite%
{Navarra:2007yw,Karliner:2017qjm,Eichten:2017ffp,Wang:2017dtg,Agaev:2019qqn}%
. The scalar tetraquark $cc\overline{u}\overline{d}$ has the mass $3845~%
\mathrm{MeV}$ higher than the threshold $3735~\mathrm{MeV}$ for strong
decay to mesons $D^{0}D^{+}$. It has the width $12.4~\mathrm{MeV}$ and may
be classified as relatively narrow resonance \cite{Agaev:2019qqn}. The
charmed partner of the axial-vector tetraquark $T_{bb}^{-}$, i.e., four
quark state $cc\overline{u}\overline{d}$ with $J^{P}=1^{+}$ was explored in
Refs.\ \cite{Navarra:2007yw,Karliner:2017qjm,Eichten:2017ffp,Wang:2017dtg}.
It turned out that mass of this state is higher than corresponding two-meson
threshold, which makes it unstable against strong decays. As expected,
tetraquarks $T_{cc}$ are unstable particles, and one of main problems is
investigation their strong decay channels. Production mechanisms of exotic
mesons $T_{cc}$ in the heavy ion and proton-proton collisions,
electron-positron annihilations, in decays of $B_{c}$ meson and heavy $\Xi
_{bc}$ baryon were addressed in the literature \cite%
{SchaffnerBielich:1998ci,DelFabbro:2004ta,Lee:2007tn,Hyodo:2012pm,Esposito:2013fma}%
.

There are exotic mesons with quark content different than ones considered
till now, which nevertheless are stable particles. One of them, namely the
scalar tetraquark $T_{bs;\overline{u}\overline{d}}^{-}$ was studied in Ref.
\cite{Agaev:2019wkk}, where we computed its mass, coupling and full width.

In the present article we are going to continue our analysis of exotic
mesons $bb\overline{u}\overline{d}$ and explore the scalar partner of $%
T_{bb}^{-}$ with the same quark content. We denote this particle $T_{b:%
\overline{d}}^{-}$ and compute its spectroscopic parameters, full width and
mean lifetime. The mass $m$ and coupling $f$ of the tetraquark $T_{b:%
\overline{d}}^{-}$ are evaluated by means of the QCD two-point sum rule
method, where we take into account various quark, gluon, and mixed vacuum
condensates up to dimension ten. A result for the mass of the state $T_{b:%
\overline{d}}^{-}$ is important for our following investigations. Indeed, $m$
determines whether $T_{b:\overline{d}}^{-}$ is strong-interaction stable
particle or not. A simple consideration allows one to see that the scalar
tetraquark $T_{b:\overline{d}}^{-}$ in $S$-wave can strongly fall-apart to a
pair of conventional mesons $B^{-}\overline{B}^{0}$ provided its mass is
higher than the threshold $10560~\mathrm{MeV}$. But, our calculations
demonstrate that mass of this tetraquark is $m=(10135\pm 240)~\mathrm{MeV}$,
and therefore $T_{b:\overline{d}}^{-}$ is a strong-interaction stable
particle. It is also stable against an electromagnetic dissociation $T_{b:%
\overline{d}}^{-}\rightarrow B^{-}\overline{B}_{1}(5721)^{0}\gamma $,
because this process may run only if the mass of the initial particle
exceeds $11003~\mathrm{MeV}$ which is not a case. As a result, to determine
the full width and mean lifetime of $T_{b:\overline{d}}^{-}$ we have to
study its weak decays.

The dominant weak decays of the $T_{b:\overline{d}}^{-}$ are generated by a
subprocess $b\rightarrow W^{-}c$ which lead to its semileptonic and
nonleptonic transformation to the exotic scalar meson $\widetilde{Z}%
_{bc}^{0}\ $(a brief form of $\widetilde{Z}_{bc;\overline{u}\overline{d}%
}^{0} $ ). We model $\widetilde{Z}_{bc}^{0}$ as a tetraquark composed of the
color-sextet diquark and antidiquark $[\mathbf{6}_{c}]_{bc}\otimes \lbrack
\overline{\mathbf{6}}_{c}]_{\overline{u}\overline{d}}$: Reasons for such
choice will be explained in the next section. The weak processes to be
explored are the semileptonic $T_{b:\overline{d}}^{-}\rightarrow \widetilde{Z%
}_{bc}^{0}l\overline{\nu }_{l}$ , and nonleptonic $T_{b:\overline{d}%
}^{-}\rightarrow \widetilde{Z}_{bc}^{0}M$ decays of $T_{b:\overline{d}}^{-}$%
. In the present work, we consider the decays, where $M$ is one of the
conventional pseudoscalar mesons $\pi ^{-}$, $K^{-}$, $D^{-}$ and $D_{s}^{-}$%
. It is clear that nonleptonic decays can be kinematically realized if $m-%
\widetilde{m}_{Z}>m_{M}$ with $\widetilde{m}_{Z}$ and $m_{M}$ being the
masses of the tetraquark $\widetilde{Z}_{bc}^{0}$ and meson $M$ ,
respectively. The spectroscopic parameters $\widetilde{m}_{Z}$, and $%
\widetilde{f}_{Z}$ of the scalar tetraquark $\widetilde{Z}_{bc}^{0}$ are
necessary to calculate partial widths of all weak decays under
consideration, and will be found as well.

The full width of the tetraquark $T_{b:\overline{d}}^{-}$ is calculated by
taking into account aforementioned semileptonic and nonleptonic decay modes.
For these purposes, we employ the QCD three-point sum rule approach and
compute weak form factors $G_{1}(q^{2})$ and $G_{2}(q^{2})$ required for our
studies. These form factors enter into differential rate $d\Gamma /dq^{2}$
of semileptonic and partial width of nonleptonic processes. The sum rule
computations, unfortunately, lead to reliable predictions for $%
G_{1(2)}(q^{2})$ only at limited values of the momentum transfers $q^{2}$.
To integrate $d\Gamma /dq^{2}$ over $m_{l}^{2}\leq q^{2}\leq (m-\widetilde{m}%
_{Z})^{2}$, and find the partial widths of the semileptonic decays, we need
to extrapolate these predictions to whole $q^{2}$ domain. The latter is
achieved by introducing fit functions $\mathcal{G}_{1(2)}(q^{2})$ that
coincide with the sum rule results when they are accessible, and can be
easily extrapolated to all $q^{2}$.

This article is organized in the following manner: In Sec.\ \ref{sec:Masses}%
, we calculate the mass and coupling of the scalar tetraquarks $T_{b:%
\overline{d}}^{-}$ and $\widetilde{Z}_{bc}^{0}$. To this end, we derive sum
rules from analysis of the relevant two-point correlation functions: in
numerical computations we take into account quark, gluon and mixed
condensates up to dimension ten. In Sec.\ \ref{sec:Decays1}, using
spectroscopic parameters of the initial and final tetraquarks and
three-point sum rules, we compute the weak form factors $G_{1(2)}(q^{2})$ in
regions of the momentum transfers $q^{2}$, where the method leads to
reliable predictions. In this section we also determine the model functions $%
\mathcal{G}_{1(2)}(q^{2})$ and find the partial widths of the semileptonic
decays\ $T_{b:\overline{d}}^{-}\rightarrow \widetilde{Z}_{bc}^{0}l\overline{%
\nu }_{l}$. In the next Sec.\ \ref{sec:Decays2} we explore the nonleptonic
decays $T_{b:\overline{d}}^{-}\rightarrow \widetilde{Z}_{bc}^{0}M$ of the
tetraquark $T_{b:\overline{d}}^{-}$. Here, we write down our final
predictions for the full width and lifetime of the $T_{b:\overline{d}}^{-}$.
Section \ref{sec:Disc} is devoted to analysis of obtained results, and
contains our concluding remarks.

%%%%%%%%%%%%%%%%%%%%%%%%%%%%%%%%%%%%%%%%%%%%%%%%%%%%%%%%%%%%%%%%%%%%%%%%%%%%

\section{Spectroscopic parameters of the scalar tetraquarks $T_{b:\overline{d%
}}^{-}$ and $\widetilde{Z}_{bc}^{0}$}

\label{sec:Masses}
%%%%%%%%%%%%%%%%%%%%%%%%%%%%%%%%%%%%%%%%%%%%%%%%%%%%%%%%%%%
The spectroscopic parameters $m$, and $f$ of the tetraquark $T_{b:\overline{d%
}}^{-}$ are required to reveal its nature and answer a question whether this
state is stable against strong and electromagnetic decays or not. The mass
and coupling of $\widetilde{Z}_{bc}^{0}$ are important to explore the weak
decays of the master particle $T_{b:\overline{d}}^{-}$. Besides, the
tetraquark $\widetilde{Z}_{bc}^{0}$, as its partner state $Z_{bc}^{0}$, may
be strong- and/or electromagnetic-interaction stable particle, which is of
independent interest for us.

The parameters of these states can be extracted from the two-point
correlation function
\begin{equation}
\Pi (p)=i\int d^{4}xe^{ipx}\langle 0|\mathcal{T}\{J(x)J^{\dag
}(0)\}|0\rangle ,  \label{eq:CF1}
\end{equation}%
where $J(x)$ is the interpolating current for a scalar particle. In the case
of $T_{b:\overline{d}}^{-}$ it is given by the expression
\begin{equation}
J(x)=[b_{a}^{T}(x)C\gamma _{5}b_{b}(x)][\overline{u}_{a}(x)\gamma _{5}C%
\overline{d}_{b}^{T}(x)],  \label{eq:CR1}
\end{equation}%
where $a$ and $b$ are color indices, and $C$ is the charge-conjugation
operator.

The current $J(x)$ is built of the color-sextet scalar diquark and
antidiquark, because the structure $[\mathbf{6}_{c}]_{bb}\otimes \lbrack
\overline{\mathbf{6}}_{c}]_{\overline{u}\overline{d}}$ is only possible
color organization for $T_{b:\overline{d}}^{-}$. In fact, the heavy diquark $%
b_{a}^{T}C\gamma _{5}b_{b}$ is composed of two $b$ quarks and has flavor
symmetric structure, therefore it should be symmetric in color indices as
well. As a result, the light antidiquark has to belong to antisextet
representation of the color group. The relevant antidiquark field $\overline{%
u}_{a}\gamma _{5}C\overline{d}_{b}^{T}+$ $\overline{u}_{b}\gamma _{5}C%
\overline{d}_{a}^{T}$ is symmetric under replacement $a\leftrightarrow b$.
But, because components of this field in the current $J(x)$ generate two
equal terms, we keep in Eq.\ (\ref{eq:CR1}) one of them.

The final-state tetraquark $bc\overline{u}\overline{d}$, in general, may be
composed of either color triplet or sextet diquaks. It is known that
four-quark mesons with scalar color-antitriplet diquark and color-triplet
antidiquark constituents are lowest lying particles, because such two-quark
structures are most attractive and stable ones \cite{Jaffe:2004ph}. Problem
is that, a matrix element for weak transition of the tetraquark $T_{b:%
\overline{d}}^{-}$ to a scalar state $bc\overline{u}\overline{d}$ with
color-triplet constituents is equal to zero identically. Therefore, we
construct the scalar particle $\widetilde{Z}_{bc}^{0}$ from color-sextet
diquarks, and choose its interpolating current in the following form
\begin{eqnarray}
\widetilde{J}_{Z}(x) &=&[b_{a}^{T}(x)C\gamma _{5}c_{b}(x)]\left[ \overline{u}%
_{a}(x)\gamma _{5}C\overline{d}_{b}^{T}(x)\right.  \notag \\
&&+\overline{u}_{b}(x)\gamma _{5}C\overline{d}_{a}^{T}(x)].  \label{eq:CR2}
\end{eqnarray}%
The color-symmetric nature of the antidiquark field in $\widetilde{J}_{Z}(x)$
is evident. The scalar diquark $bc$ can be interpolated by the field $%
b_{a}^{T}C\gamma _{5}c_{b}+b_{b}^{T}C\gamma _{5}c_{a}$, but its components
lead to equal currents, as a result, in Eq.\ (\ref{eq:CR2}) we use one of
these terms.
\begin{figure}[h]
\includegraphics[width=8.8cm]{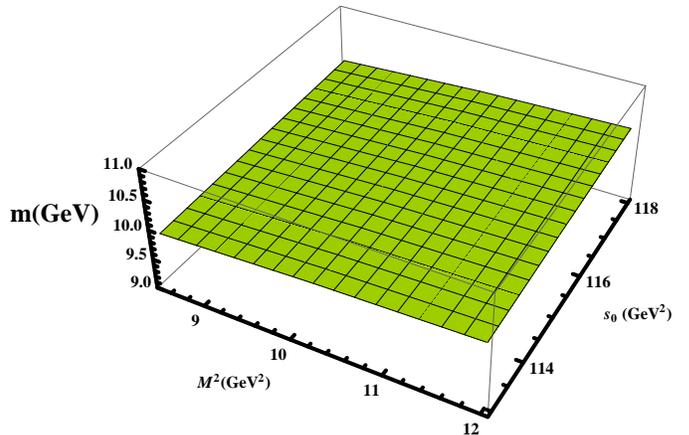}
\caption{The mass $m$ of the tetraquark $T_{b:\overline{d}^{-}}$ as a
function of the Borel $M^{2}$ and continuum threshold $s_{0}$ parameters.}
\label{fig:Mass}
\end{figure}

Spectroscopic parameters of different scalar tetraquarks were objects of
detailed sum rule analysis, therefore we provide below only essential stages
of calculations in the case of the tetraquark $T_{b:\overline{d}}^{-}$, and
give final results for the $\widetilde{Z}_{bc}^{0}$.

The sum rules to evaluate $m$ and $f$ can be obtained by matching two
expressions of the correlation function $\Pi (p)$: the first expression is
calculated using the physical parameters of $T_{b:\overline{d}}^{-}$,
whereas the second one is written down in terms of quark propagators. The
physical side of the sum rules $\Pi ^{\mathrm{Phys}}(p)$ in the
"ground-state + continuum" scheme is given by the formula
\begin{equation}
\Pi ^{\mathrm{Phys}}(p)=\frac{\langle 0|J|T_{b:\overline{d}}^{-}(p)\rangle
\langle T_{b:\overline{d}}^{-}(p)|J^{\dagger }|0\rangle }{m^{2}-p^{2}}%
+\cdots.  \label{eq:Phen1}
\end{equation}%
In Eq.\ (\ref{eq:Phen1}) we write down contribution of only the ground-state
tetraquark, and denote by dots effects of higher resonances and continuum
states. Using definition of the spectroscopic parameters of $T_{b:\overline{d%
}}^{-}$ through the matrix element
\begin{equation}
\langle 0|J|T_{b:\overline{d}}^{-}(p)\rangle =fm,  \label{eq:ME1}
\end{equation}%
we recast $\Pi ^{\mathrm{Phys}}(p)$ into the final form%
\begin{equation}
\Pi ^{\mathrm{Phys}}(p)=\frac{f^{2}m^{2}}{m^{2}-p^{2}}+\cdots .
\label{eq:Phen2}
\end{equation}%
The function $\Pi ^{\mathrm{Phys}}(p)$ has a trivial Lorentz structure
proportional to $\sim I$, and the term in Eq.\ (\ref{eq:Phen2}) is the
invariant amplitude $\Pi ^{\mathrm{Phys}}(p^{2})$ corresponding to this
structure.

To determine the second component of the sum rule analysis, we calculate $%
\Pi (p)$ using the quark-gluon degrees of freedom. For these purposes, we
insert the explicit expression of the interpolating current $J(x)$ into Eq.\
(\ref{eq:Phen1}), and contract relevant heavy and light quark fields. After
these operations for $\Pi ^{\mathrm{OPE}}(p)$ we get
\begin{eqnarray}
&&\Pi ^{\mathrm{OPE}}(p)=i\int d^{4}xe^{ipx}\mathrm{Tr}\left[ \gamma _{5}%
\widetilde{S}_{d}^{b^{\prime }b}(-x)\gamma _{5}S_{u}^{a^{\prime }a}(-x)%
\right]  \notag \\
\times &&\left\{ \mathrm{Tr}\left[ \gamma _{5}\widetilde{S}_{b}^{aa^{\prime
}}(x)\gamma _{5}S_{b}^{bb^{\prime }}(x)\right] +\mathrm{Tr}\left[ \gamma _{5}%
\widetilde{S}_{b}^{ba^{\prime }}(x)\gamma _{5}S_{b}^{ab^{\prime }}(x)\right]
\right\} ,  \notag \\
&&  \label{eq:QCD1}
\end{eqnarray}%
where $S_{b}(x)$ and $S_{u(d)}(x)$ are the heavy $b$- and light $u(d)$-quark
propagators, respectively. Above we also introduce the shorthand notation
\begin{equation}
\widetilde{S}_{b(u,d)}(x)=CS_{b(u,d)}^{T}(x)C.  \label{eq:Notation}
\end{equation}%
The explicit expressions of the heavy and light quark propagators can be
found, for instance, in Ref.\ \cite{Agaev:2020zad}. The nonperturbative
parts of the propagators contain various quark, gluon, and mixed condensates
which are sources of nonperturbative terms in $\Pi ^{\mathrm{OPE}}(p)$.

The first equality necessary to derive the sum rules are obtained by
equating the amplitudes $\Pi ^{\mathrm{Phys}}(p^{2})$ and $\Pi ^{\mathrm{OPE}%
}(p^{2})$, and applying to both sides of this expression the Borel
transformation: By this way we suppress contributions to the sum rules of
higher resonances and continuum states. But even after the Borel
transformation suppressed terms appear as a contamination in the physical
side of the equality. Fortunately, they can be subtracted by invoking
assumption about quark-hadron duality. The second equality required for our
purposes is derived by applying the operator $d/d(-1/M^{2})$ to the first
one. These two expressions are enough to get the sum rules for $m$
\begin{equation}
m^{2}=\frac{\int_{4m_{b}^{2}}^{s_{0}}dss\rho ^{\mathrm{OPE}}(s)e^{-s/M^{2}}}{%
\int_{4m_{b}^{2}}^{s_{0}}ds\rho ^{\mathrm{OPE}}(s)e^{-s/M^{2}}},
\label{eq:Mass}
\end{equation}%
and for $f$%
\begin{equation}
f^{2}=\frac{1}{m^{2}}\int_{4m_{b}^{2}}^{s_{0}}ds\rho ^{\mathrm{OPE}%
}(s)e^{(m^{2}-s)/M^{2}}.  \label{eq:Coupl}
\end{equation}%
The two-point spectral density $\rho ^{\mathrm{OPE}}(s)$ is computed as an
imaginary part of the correlation function $\Pi ^{\mathrm{OPE}}(p)$. We
include into analysis vacuum condensates up to dimension 10: because the
final expression of $\rho ^{\mathrm{OPE}}(s)$ is rather lengthy, we do not
write down it here.

The sum rules (\ref{eq:Mass}) and (\ref{eq:Coupl}) contain the universal
vacuum condensates and masses of $b$ and $c$ quarks:%
\begin{eqnarray}
&&\langle \bar{q}q\rangle =-(0.24\pm 0.01)^{3}~\mathrm{GeV}^{3},\ \langle
\bar{s}s\rangle =0.8\ \langle \bar{q}q\rangle ,  \notag \\
&&\langle \overline{q}g_{s}\sigma Gq\rangle =m_{0}^{2}\langle \overline{q}%
q\rangle ,\ \langle \overline{s}g_{s}\sigma Gs\rangle =m_{0}^{2}\langle \bar{%
s}s\rangle ,  \notag \\
&&m_{0}^{2}=(0.8\pm 0.1)~\mathrm{GeV}^{2}  \notag \\
&&\langle \frac{\alpha _{s}G^{2}}{\pi }\rangle =(0.012\pm 0.004)~\mathrm{GeV}%
^{4},  \notag \\
&&\langle g_{s}^{3}G^{3}\rangle =(0.57\pm 0.29)~\mathrm{GeV}^{6},  \notag \\
&&m_{c}=1.27\pm 0.2~\mathrm{GeV},\ m_{b}=4.18_{-0.02}^{+0.03}~\mathrm{GeV}.
\label{eq:Parameters}
\end{eqnarray}

Besides, $m$ and $f$ depend on the Borel $M^{2}$ and continuum threshold $%
s_{0}$ parameters appeared in Eqs.\ (\ref{eq:Mass}) and (\ref{eq:Coupl})
after the Borel transformation and continuum subtraction procedures,
respectively. The $M^{2}$ and $s_{0}$ are the auxiliary parameters of the
problem under discussion, a correct choice of which is an important task of
computations. But proper regions for $M^{2}$ and $s_{0}$ should meet some
restrictions imposed on the pole contribution ($\mathrm{PC}$) and
convergence of the operator product expansion ($\mathrm{OPE}$). In fact, at
maximum of $M^{2}$ the $\mathrm{PC}$ should obey the constraint
\begin{equation}
\mathrm{PC}=\frac{\Pi (M^{2},s_{0})}{\Pi (M^{2},\infty )}>0.2,  \label{eq:PC}
\end{equation}%
where $\Pi (M^{2},s_{0})$ is the Borel-transformed and subtracted invariant
amplitude $\Pi ^{\mathrm{OPE}}(p^{2})$. The minimum of $M^{2}$ is fixed \
from analysis of the ratio
\begin{equation}
R(M^{2})=\frac{\Pi ^{\mathrm{DimN}}(M^{2},s_{0})}{\Pi (M^{2},s_{0})}\leq
0.01.  \label{eq:Convergence}
\end{equation}%
In Eq.\ (\ref{eq:Convergence}) $\Pi ^{\mathrm{DimN}}(M^{2},s_{0})$ denotes a
contribution of the last term (or a sum of last few terms) to the
correlation function. In the present calculations we use the sum of last
three terms, and hence $\mathrm{DimN\equiv Dim(8+9+10)}$.

Our analysis demonstrates that the working windows for the parameters $M^{2}$
and $s_{0}$ are
\begin{equation}
M^{2}\in \lbrack 8.5,12]\ \mathrm{GeV}^{2},\ s_{0}\in \lbrack 113,118]\
\mathrm{GeV}^{2},  \label{eq:Wind1}
\end{equation}%
and they satisfy all aforementioned constraints on $M^{2}$ and $s_{0}$.
Indeed, at $M^{2}=12~\mathrm{GeV}^{2}$ the pole contribution is $0.21$,
whereas at $M^{2}=8.5~\mathrm{GeV}^{2}$ it amounts to $0.61$. These two
values of $M^{2}$ fix the boundaries of a region where the Borel parameter
can be varied. Relatively wide range of $M^{2}$ allows us to explore the
stability of obtained predictions for $m$ and $f$. It is worth emphasizing
that, we extract these parameters approximately at a middle region of the
window (\ref{eq:Wind1}), where the pole contribution is $\mathrm{PC}\approx
0.48-0.51$. \ This fact confirms the ground state nature of the tetraquark $%
T_{b:\overline{d}}^{-}$. \ At the minimum of $M^{2}=8.5~\mathrm{GeV}^{2}$ we
get $R\approx 0.006$. Apart from that, at minimum of the Borel parameter the
perturbative contribution forms $79\%$ of the whole result overshooting
significantly the nonperturbative terms.

Our results for $m$ and $f$ are
\begin{eqnarray}
m &=&(10135~\pm 240)~\mathrm{MeV},  \notag \\
f &=&(2.26\pm 0.57)\times 10^{-2}~\mathrm{GeV}^{4},  \label{eq:Result1}
\end{eqnarray}%
where uncertainties of computations are shown as well. Theoretical
uncertainties in the case of $m$ equal to $\pm 2.4\%$, whereas for the
coupling $f$ they amount to $\pm 25\%$ remaining, at the same time, within
limits accepted in sum rule computations. It is worth noting that these
uncertainties appear mainly due to variations of the parameters $M^{2}$ and $%
s_{0}$. In Fig.\ \ref{fig:Mass} we display the sum rule's prediction for $m$
as a function of $M^{2}$ and $s_{0}$, where one can see residual dependence
of the mass on these parameters.

The mass and coupling of the scalar tetraquark $\widetilde{Z}_{bc}^{0}$ are
calculated by the same way. The phenomenological side of the corresponding
sum rules is determined by Eq.\ (\ref{eq:Phen2}) with evident replacement $%
(m,f)\rightarrow (\widetilde{m}_{Z},\widetilde{f}_{Z})$. Their QCD side is
given by the following formula
\begin{eqnarray}
&&\widetilde{\Pi }^{\mathrm{OPE}}(p)=i\int d^{4}xe^{ipx}\mathrm{Tr}\left[
\gamma _{5}\widetilde{S}_{b}^{aa^{\prime }}(x)\gamma _{5}S_{c}^{bb^{\prime
}}(x)\right]  \notag \\
&&\times \left\{ \mathrm{Tr}\left[ \gamma _{5}\widetilde{S}_{d}^{b^{\prime
}b}(-x)\gamma _{5}S_{u}^{a^{\prime }a}(-x)\right] +\mathrm{Tr}\left[ \gamma
_{5}\widetilde{S}_{d}^{a^{\prime }b}(-x)\right. \right.  \notag \\
&&\left. \times \gamma _{5}S_{u}^{b^{\prime }a}(-x)\right] +\mathrm{Tr}\left[
\gamma _{5}\widetilde{S}_{d}^{b^{\prime }a}(-x)\gamma _{5}S_{u}^{a^{\prime
}b}(-x)\right]  \notag \\
&&\left. +\mathrm{Tr}\left[ \gamma _{5}\widetilde{S}_{d}^{a^{\prime
}a}(-x)\gamma _{5}S_{u}^{b^{\prime }b}(-x)\right] \right\} .  \label{eq;OPE2}
\end{eqnarray}%
The mass $\widetilde{m}_{Z}$ and coupling $\widetilde{f}_{Z}$ of the
tetraquark $\widetilde{Z}_{bc}^{0}$ can be found from Eqs.\ (\ref{eq:Mass})
and (\ref{eq:Coupl}) by replacing $\rho ^{\mathrm{OPE}}(s)\rightarrow
\widetilde{\rho }^{\mathrm{OPE}}(s)$, where the spectral density $\widetilde{%
\rho }^{\mathrm{OPE}}(s)$ is found using the correlation function $%
\widetilde{\Pi }^{\mathrm{OPE}}(p)$, and substituting $(m_{b}+m_{c})^{2}$
instead of $4m_{b}^{2}$. As working windows for $M^{2}$ and $s_{0}$ we
utilize%
\begin{equation}
M^{2}\in \lbrack 5.5,6.5]~\mathrm{GeV}^{2},\ s_{0}\in \lbrack 53,54]~\mathrm{%
GeV}^{2}.  \label{eq:Wind2}
\end{equation}%
The regions (\ref{eq:Wind2}) obey standard constraints of the sum rule
computations. Thus, at $M^{2}=5.5~\mathrm{GeV}^{2}$ the ratio $R$ is $0.01$,
hence the convergence of the sum rules is satisfied. The pole contribution $%
\mathrm{PC}$ at $M^{2}=6.5~\mathrm{GeV}^{2}$ and $M^{2}=5.5~\mathrm{GeV}^{2}$
equals to $0.24$ and $0.71$, respectively. At minimum of $M^{2}$ the
perturbative contribution constitutes $72\%$ of the whole result exceeding
considerably nonperturbative terms.

\begin{figure}[h]
\includegraphics[width=8.8cm]{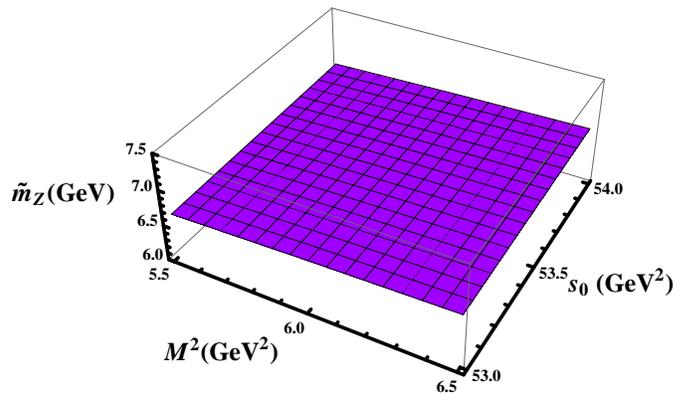}
\caption{The mass $\widetilde{m}_{Z}$ of the tetraquark $\widetilde{Z}%
_{bc}^{0}$ as a function of the parameters $M^{2}$ and $s_{0}$.}
\label{fig:MassZ}
\end{figure}

For $\widetilde{m}_{Z}$ and $\widetilde{f}_{Z}$ our computations yield
\begin{eqnarray}
\widetilde{m}_{Z} &=&(6730~\pm 150)~\mathrm{MeV},  \notag \\
\widetilde{f}_{Z} &=&(6.2\pm 1.4)\times 10^{-3}~\mathrm{GeV}^{4}.
\label{eq:Result2}
\end{eqnarray}%
In Fig.\ \ref{fig:MassZ} we plot the prediction obtained for the mass of the
tetraquark $\widetilde{Z}_{bc}^{0}$ and show its dependence on $M^{2}$ and $%
s_{0}$.

%%%%%%%%%%%%%%%%%%%%%%%%%%%%%%%%%%%%%%%%%%%%%%%%%%%%%%%%%%%%%%%%%%%%%

\section{Semileptonic decays $T_{b:\overline{d}}^{-}\rightarrow \widetilde{Z}%
_{bc}^{0}l\overline{\protect\nu }_{l}$}

\label{sec:Decays1}
%%%%%%%%%%%%%%%%%%%%%%%%%%%%%%%%%%%%%%%%%%%%%%%%%%%%%%%%%%%%%%%%%%%%%%%%%%%%%%
The result for the mass of the tetraquark $T_{b:\overline{d}}^{-}$ proves
its stability against the strong and radiative decays. In fact, the central
value of the mass $m=10135~\mathrm{MeV}$ is $425~\mathrm{MeV}$ lower than
the threshold $10560~\mathrm{MeV}$ for strong decay to mesons $B^{-}%
\overline{B}^{0}$. Its maximal allowed value $m_{\mathrm{max}}=10375~\mathrm{%
MeV}$ is $185~\mathrm{MeV}$ below this limit as well. In other words, the $%
T_{b:\overline{d}}^{-}$ is a strong-interaction stable particle. The
threshold $11003~\mathrm{MeV}$ for the decay $T_{b:\overline{d}%
}^{-}\rightarrow B^{-}\overline{B}_{1}(5721)^{0}\gamma $ is higher than $m_{%
\mathrm{max}}$ which forbids this electromagnetic process. Therefore, the
full width and mean lifetime of the $T_{b:\overline{d}}^{-}$ are determined
by its weak decays.

This section is devoted to analysis of the dominant semileptonic decay $T_{b:%
\overline{d}}^{-}\rightarrow \widetilde{Z}_{bc}^{0}l\overline{\nu }_{l}$
triggered by the weak transition of the heavy $b$-quark $b\rightarrow
W^{-}c\rightarrow cl\overline{\nu }$. It is evident, that the mass
difference\ $3405~\mathrm{MeV}$ between the states $T_{b:\overline{d}}^{-}$
and $\widetilde{Z}_{bc}^{0}$ makes all decays $T_{b:\overline{d}%
}^{-}\rightarrow \widetilde{Z}_{bc}^{0}l\overline{\nu }_{l}$, where $l=e,\
\mu $ and $\tau $ kinematically allowed ones. Here, we neglect processes
generated by a subprocess $b\rightarrow W^{-}u$, because they are suppressed
relative to dominant decays by a factor $|V_{bu}|^{2}/|V_{bc}|^{2}$ $\simeq
0.01$ with $V_{q_{1}q_{2}}$ being the Cabibbo-Khobayasi-Maskawa (CKM) matrix
elements.

At the tree-level the subprocess $b\rightarrow W^{-}c$ can be described
using the effective Hamiltonian
\begin{equation}
\mathcal{H}^{\mathrm{eff}}=\frac{G_{F}}{\sqrt{2}}V_{bc}\overline{c}\gamma
_{\mu }(1-\gamma _{5})b\overline{l}\gamma ^{\mu }(1-\gamma _{5})\nu _{l}.
\label{eq:EffecH}
\end{equation}%
Here, $G_{F}$ and $V_{bc}$ are the Fermi coupling constant and CKM matrix
element, respectively%
\begin{eqnarray}
G_{F} &=&1.16637\times 10^{-5}~\mathrm{GeV}^{-2},  \notag \\
|V_{bc}| &=&(42.2\pm 0.08)\times 10^{-3}.
\end{eqnarray}%
A matrix element of $\mathcal{H}^{\mathrm{eff}}$ between the initial and
final tetraquarks
\begin{equation}
\langle \widetilde{Z}_{bc}^{0}(p^{\prime })|\mathcal{H}^{\mathrm{eff}}|T_{b:%
\overline{d}}^{-}(p)\rangle =L_{\mu }H^{\mu },
\end{equation}%
consists of leptonic and hadronic factors. A leptonic part of the matrix
element $L_{\mu }$ is universal for all semileptonic decays and does not
contain information on features of tetraquarks. Therefore, we are interested
in calculation of $H^{\mu }$ which is nothing more than the matrix element
of the current
\begin{equation}
J_{\mu }^{\mathrm{tr}}=\overline{c}\gamma _{\mu }(1-\gamma _{5})b.
\label{eq:TrCurr}
\end{equation}%
It can be detailed using form factors $G_{1(2)}(q^{2})$ that parametrize the
long-distance dynamics of the weak transition. In terms of $G_{1(2)}(q^{2})$
the matrix element $H^{\mu }$ has the form
\begin{eqnarray}
\langle \widetilde{Z}_{bc}^{0}(p^{\prime })|J_{\mu }^{\mathrm{tr}}|T_{b:%
\overline{d}}^{-}(p)\rangle &=&G_{1}(q^{2})P_{\mu }+G_{2}(q^{2})q_{\mu },
\notag \\
&&  \label{eq:Vertex1}
\end{eqnarray}%
where $p$ and $p^{\prime }$ are the momenta of the initial and final
tetraquarks, respectively. Above, we also use notations $P_{\mu }=p_{\mu
}^{\prime }+p_{\mu }$ and $q_{\mu }=p_{\mu }-p_{\mu }^{\prime }$. The $%
q_{\mu }$ is the momentum transferred to the leptons, hence $q^{2}$ changes
in the region $m_{l}^{2}\leq q^{2}\leq (m-\widetilde{m}_{Z})^{2}$, where $%
m_{l}$ is the mass of a lepton $l$.

The sum rules for the form factors $G_{1(2)}(q^{2})$ can be extracted from
the three-point correlation function
\begin{eqnarray}
\Pi _{\mu }(p,p^{\prime }) &=&i^{2}\int d^{4}xd^{4}ye^{i(p^{\prime }y-px)}
\notag \\
&&\times \langle 0|\mathcal{T}\{\widetilde{J}_{Z}(y)J_{\mu }^{\mathrm{tr}%
}(0)J^{\dagger }(x)\}|0\rangle .  \label{eq:CF2}
\end{eqnarray}

As usual, we write down $\Pi _{\mu }(p,p^{\prime })$ using the spectroscopic
parameters of the tetraquarks, and get the physical side of the sum rule $%
\Pi _{\mu }^{\mathrm{Phys}}(p,p^{\prime })$. \ The function $\Pi _{\mu }^{%
\mathrm{Phys}}(p,p^{\prime })$ has the following form%
\begin{eqnarray}
&&\Pi _{\mu }^{\mathrm{Phys}}(p,p^{\prime })=\frac{\langle 0|\widetilde{J}%
_{Z}|\widetilde{Z}_{bc}^{0}(p^{\prime })\rangle \langle \widetilde{Z}%
_{bc}^{0}(p^{\prime })|J_{\mu }^{\mathrm{tr}}|T_{b:\overline{d}%
}^{-}(p)\rangle }{(p^{2}-m^{2})(p^{\prime 2}-\widetilde{m}_{Z}^{2})}  \notag
\\
&&\times \langle T_{b:\overline{d}}^{-}(p)|J^{\dagger }|0\rangle +\cdots ,
\label{eq:CF3}
\end{eqnarray}%
where the term in Eq.\ (\ref{eq:CF3}) is contribution of the ground-state
particles: contributions of excited resonances and continuum states are
denoted by dots.

The phenomenological side of the sum rules can be simplified by substituting
in Eq.\ (\ref{eq:CF3}) expressions of matrix elements in terms of the
tetraquarks' masses and couplings, and weak transition form factors. To this
end, we employ Eqs.\ (\ref{eq:ME1}) and (\ref{eq:Vertex1}), and additionally
invoke the matrix element of the state $\widetilde{Z}_{bc}^{0}$
\begin{equation}
\langle 0|\widetilde{J}_{Z}|\widetilde{Z}_{bc}^{0}(p^{\prime })\rangle =%
\widetilde{f}_{Z}\widetilde{m}_{Z}.  \label{eq:ME3}
\end{equation}%
Then one gets
\begin{eqnarray}
\Pi _{\mu }^{\mathrm{Phys}}(p,p^{\prime }) &=&\frac{fm\widetilde{f}_{Z}%
\widetilde{m}_{Z}}{(p^{2}-m^{2})(p^{\prime 2}-\widetilde{m}_{Z}^{2})}  \notag
\\
&&\times \left[ G_{1}(q^{2})P_{\mu }+G_{2}(q^{2})q_{\mu }\right] +\cdots .
\label{eq:Phys1}
\end{eqnarray}

We find $\Pi _{\mu }(p,p^{\prime })$ also using explicitly the interpolating
currents in the correlator, and expressing (\ref{eq:CF2}) in terms of quark
propagators, which lead to the QCD side of the sum rules
\begin{eqnarray}
&&\Pi _{\mu }^{\mathrm{OPE}}(p,p^{\prime })=i^{2}\int
d^{4}xd^{4}ye^{i(p^{\prime }y-px)}\left( \mathrm{Tr}\left[ \gamma _{5}%
\widetilde{S}_{d}^{b^{\prime }b}(x-y)\right. \right.  \notag \\
&&\left. \left. \times \gamma _{5}S_{u}^{a^{\prime }a}(x-y)\right] +\mathrm{%
Tr}\left[ \gamma _{5}\widetilde{S}_{d}^{b^{\prime }a}(x-y)\gamma _{\nu
}S_{u}^{a^{\prime }b}(x-y)\right] \right)  \notag \\
&&\times \left( \mathrm{Tr}\left[ \gamma _{5}\widetilde{S}_{b}^{aa^{\prime
}}(y-x)\gamma _{5}S_{c}^{bi}(y)\gamma _{\mu }(1-\gamma
_{5})S_{b}^{ib^{\prime }}(y)\right] \right.  \notag \\
&&\left. +\mathrm{Tr}\left[ \gamma _{5}\widetilde{S}_{b}^{ia^{\prime
}}(-x)(1-\gamma _{5})\gamma _{\mu }\widetilde{S}_{c}^{bi}(y)\gamma
_{5}S_{b}^{ab^{\prime }}(y-x)\right] \right) .  \notag \\
&&  \label{eq:QCD2}
\end{eqnarray}

It is seen that the correlator $\Pi _{\mu }(p,p^{\prime })$ has structures
proportional to $P_{\mu }$ and $q_{\mu }$. Extracting from $\Pi _{\mu }^{%
\mathrm{Phys}}(p,p^{\prime })$ and $\Pi _{\mu }^{\mathrm{OPE}}(p,p^{\prime
}) $ invariant amplitudes corresponding to these structures, and equating
them to each other, we can derive sum rules for the form factors $%
G_{1(2)}(q^{2})$. One of the main procedures in our computations is the
Borel transformation of obtained equalities. Because relevant amplitudes
depend on $p^{2}$ and $p^{\prime 2}$, in order to suppress contributions of
higher resonances and continuum states, we should apply the double Borel
transformation over these variables. Final expressions obtained after these
operations depend on a set of Borel parameters $\mathbf{M}^{2}=(M_{1}^{2},\
M_{2}^{2})$. Then the continuum subtraction should also be carried out in
two channels by introducing a set of threshold parameters $\mathbf{s}%
_{0}=(s_{0},\ s_{0}^{\prime })$.

After these manipulations, we derive the sum rules
\begin{eqnarray}
&&G_{i}(\mathbf{M}^{2},\mathbf{s}_{0},q^{2})=\frac{1}{fm\widetilde{f}_{Z}%
\widetilde{m}_{Z}}\int_{4m_{b}^{2}}^{s_{0}}dse^{(m^{2}-s)/M_{1}^{2}}  \notag
\\
&&\times \int_{(m_{b}+m_{c})^{2}}^{s_{0}^{\prime }}ds^{\prime }\rho
_{i}(s,s^{\prime },q^{2})e^{(\widetilde{m}_{Z}^{2}-s^{\prime })/M_{2}^{2}},
\label{eq:SR}
\end{eqnarray}%
where $\rho _{1(2)}(s,s^{\prime },q^{2})$ are the spectral densities
calculated with dimension-7 accuracy. In Eq.\ (\ref{eq:SR}) the pair of
parameters $(M_{1}^{2},s_{0})$ describes the initial tetraquark $T_{b:%
\overline{d}}^{-}$, whereas the set $(M_{2}^{2},s_{0}^{\prime })$
corresponds to the final state $\widetilde{Z}_{bc}^{0}$.

\begin{figure}[h]
\includegraphics[width=8.8cm]{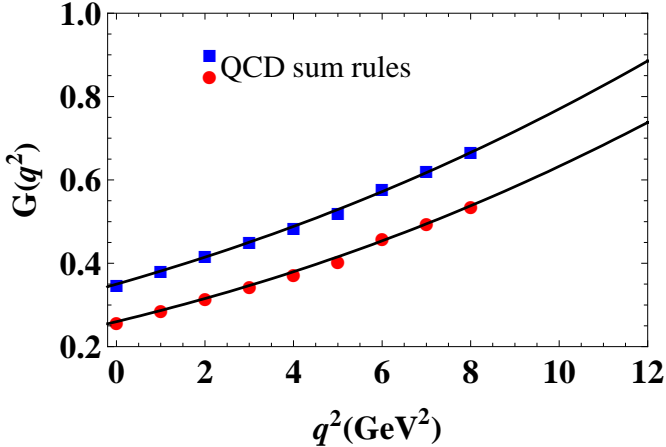}
\caption{Predictions for the form factors $|G_{1}(q^{2})|$ (the lower red
circles) and $G_{2}(q^{2})$ (the upper blue squares). The lines are the fit
functions $|\mathcal{G}_{1}(q^{2})|$ and $\mathcal{G}_{2}(q^{2})$,
respectively.}
\label{fig:WFF}
\end{figure}

In computations the working regions for $\mathbf{M}^{2}$ and $\mathbf{s}_{0}
$ are chosen as in analyses of the masses $m$ and $\widetilde{m}_{Z}$. Input
information necessary for numerical calculations of $G_{1(2)}(q^{2})$ that
includes the vacuum condensates, spectroscopic parameters of the tetraquarks
$T_{b:\overline{d}}^{-}$ and $\widetilde{Z}_{bc}^{0}$ are presented in Eqs.\
(\ref{eq:Parameters}), \ (\ref{eq:Result1}) and (\ref{eq:Result2}),
respectively. In Fig.\ \ref{fig:WFF} we show obtained predictions for the
form factors $G_{1}(q^{2})$ and $G_{2}(q^{2})$.

The sum rules give reliable results for $G_{1(2)}(q^{2})$ in the region $%
m_{l}^{2}\leq q^{2}\leq 8~\mathrm{GeV}^{2}$, which is not enough to
calculate the partial width of the process $T_{b:\overline{d}%
}^{-}\rightarrow \widetilde{Z}_{bc}^{0}l\overline{\nu }_{l}$ under analysis.
Thus, the form factors $G_{1(2)}(q^{2})$ determine the differential decay
rate $d\Gamma /dq^{2}$ of this process%
\begin{eqnarray}
&&\frac{d\Gamma }{dq^{2}}=\frac{G_{F}^{2}|V_{bc}|^{2}}{64\pi ^{3}m^{3}}%
\lambda \left( m^{2},\widetilde{m}_{Z}^{2},q^{2}\right) \left( \frac{%
q^{2}-m_{l}^{2}}{q^{2}}\right) ^{2}  \notag \\
\times &&\left\{ (2q^{2}+m_{l}^{2})\left[ G_{1}^{2}(q^{2})\left( \frac{q^{2}%
}{2}-m^{2}-\widetilde{m}_{Z}^{2}\right) \right. \right.  \notag \\
&&\left. -G_{2}^{2}(q^{2})\frac{q^{2}}{2}+(\widetilde{m}%
_{Z}^{2}-m^{2})G_{1}(q^{2})G_{2}(q^{2})\right]  \notag \\
&&\left. +\frac{q^{2}+m_{l}^{2}}{q^{2}}\left[ G_{1}(q^{2})(m^{2}-\widetilde{m%
}_{Z}^{2})+G_{2}(q^{2})q^{2}\right] ^{2}\right\} ,  \notag \\
&&  \label{eq:Drate}
\end{eqnarray}%
where%
\begin{eqnarray}
&&\lambda \left( m^{2},\widetilde{m}_{Z}^{2},q^{2}\right) =\left[ m^{4}+%
\widetilde{m}_{Z}^{4}+q^{4}\right.  \notag \\
&&\left. -2\left( m^{2}\widetilde{m}_{Z}^{2}+m^{2}q^{2}+\widetilde{m}%
_{Z}^{2}q^{2}\right) \right] ^{1/2}.  \label{eq:Lambda}
\end{eqnarray}%
To find the width of a semileptonic decay, $d\Gamma /dq^{2}$ should be
integrated over $q^{2}$ in the limits $m_{l}^{2}\leq q^{2}\leq (m-\widetilde{%
m}_{Z})^{2}$. But $m_{l}^{2}\leq q^{2}\leq 11.59~\mathrm{GeV}^{2}$ is wider
than the region where the sum rules lead to strong results. This problem can
be evaded by introducing fit functions $\mathcal{G}_{i}(q^{2})$ ($i=1,2$):
at the momentum transfers $q^{2}$ accessible for the sum rule computations
they have to coincide with $G_{i}(q^{2})$, but have analytic forms suitable
to carry out integrations over $q^{2}$.

For these purposes, we use the functions of the form
\begin{equation}
\mathcal{G}_{i}(q^{2})=\mathcal{G}_{0}^{i}\exp \left[ g_{1}^{i}\frac{q^{2}}{%
m^{2}}+g_{2}^{i}\left( \frac{q^{2}}{m^{2}}\right) ^{2}\right] ,
\label{eq:FFunctions}
\end{equation}%
where $\mathcal{G}_{0}^{i},~g_{1}^{i},\ $ and $g_{2}^{i}$ are constants
which have to be fixed by comparing $\mathcal{G}_{i}(q^{2})$ and $%
G_{i}(q^{2})$ at common domains of validity. Performed numerical analysis
gives
\begin{eqnarray}
\mathcal{G}_{0}^{1} &=&-0.26,~g_{1}^{1}=10.14,\ g_{2}^{1}=-10.36  \notag \\
\mathcal{G}_{0}^{2} &=&0.35,~g_{1}^{2}=8.87,\ g_{2}^{2}=-7.91.
\label{eq:Fitparameters}
\end{eqnarray}%
The functions $\mathcal{G}_{i}(q^{2})$ are plotted in Fig. \ref{fig:WFF}:
one can see an agreement between the sum rule predictions and fit functions.

The masses of the leptons $m_{e}=0.511~\mathrm{MeV}$, $m_{\mu }=105.658~%
\mathrm{MeV}$, and $m_{\tau }=(1776.82~\pm 0.16)~\mathrm{MeV}$ used to find $%
\Gamma (T_{b:\overline{d}}^{-}\rightarrow \widetilde{Z}_{bc}^{0}l\overline{%
\nu }_{l})$ are taken from Ref.\ \cite{Tanabashi:2018oca}. The results
obtained for the partial width of the semileptonic decays $T_{b:\overline{d}%
}^{-}\rightarrow \widetilde{Z}_{bc}^{0}l\overline{\nu }_{l}$ are collected
in Table \ref{tab:PWidth}.
\begin{table}[tbp]
\begin{tabular}{|c|c|}
\hline\hline
Channel & Partial width \\ \hline\hline
$T_{b:\overline{d}}^{-} \to \widetilde{Z}_{bc}^{0}e^{-}\overline{\nu }_{e}$
& $(4.45 \pm 1.28)\times 10^{-10}~\mathrm{MeV}$ \\
$T_{b:\overline{d}}^{-} \to \widetilde{Z}_{bc}^{0}\mu ^{-}\overline{\nu }%
_{\mu }$ & $(4.44\pm 1.26)\times 10^{-10}~\mathrm{MeV}$ \\
$T_{b:\overline{d}}^{-} \to \widetilde{Z}_{bc}^{0}\tau ^{-}\overline{\nu }%
_{\tau }$ & $(1.99\pm 0.56)\times 10^{-10}~\mathrm{MeV}$ \\
$T_{b:\overline{d}}^{-} \to \widetilde{Z}_{bc}^{0}\pi ^{-}$ & $(5.13\pm
1.42)\times 10^{-13}~\mathrm{MeV}$ \\
$T_{b:\overline{d}}^{-} \to \widetilde{Z}_{bc}^{0}K^{-}$ & $(3.93\pm
1.12)\times 10^{-14}~\mathrm{MeV}$ \\
$T_{b:\overline{d}}^{-} \to \widetilde{Z}_{bc}^{0}D^{-}$ & $(8.49\pm
2.41)\times 10^{-14}~\mathrm{MeV}$ \\
$T_{b:\overline{d}}^{-} \to \widetilde{Z}_{bc}^{0}D_{s}^{-}$ & $(2.92\pm
0.82)\times 10^{-12}~\mathrm{MeV}$ \\ \hline\hline
\end{tabular}%
\caption{Partial width of the $T_{b:\overline{d}}^{-}$ tetraquark's weak
decay channels.}
\label{tab:PWidth}
\end{table}

%%%%%%%%%%%%%%%%%%%%%%%%%%%%%%%%%%%%%%%%%%%%%%%%%%%%%%%%%%%%%

\section{ Nonleptonic decays $T_{b:\overline{d}}^{-}\rightarrow \widetilde{Z}%
_{bc}^{0}M$}

\label{sec:Decays2}
%%%%%%%%%%%%%%%%%%%%%%%%%%%%%%%%%%%%%%%%%%%%%%%%%%%%%%%%%%%%%%%%%%%%

Nonleptonic decays of $T_{b:\overline{d}}^{-}$ may be generated by weak
transformations of constituent quarks (antiquarks) of $T_{b:\overline{d}%
}^{-} $ provided these processes are kinematically allowed. The subprocesses
$\overline{u}\rightarrow \overline{d}W^{-}$,\ $\overline{u}\rightarrow
\overline{s}W^{-}$ and $\overline{u}\rightarrow \overline{d}W^{-}$ imply
production of tetraquarks $bb\overline{d}\overline{d}$, $bb\overline{s}%
\overline{d}$ and $bb\overline{b}\overline{d}$, respectively, and a meson.
It is clear that such processes are forbidden kinematically, because the
mass of a produced tetraquark is either equal to or higher than the mass $m$
of $T_{b:\overline{d}}^{-}$ (in the present work $m_{u}=m_{d}\equiv 0$). The
same arguments are true also for weak transitions of the antidiquark $%
\overline{d}$. The dominant nonleptonic decays of $T_{b:\overline{d}}^{-}$
is triggered by the subprocess $b\rightarrow W^{-}c$, whereas the transition
$b\rightarrow W^{-}u$ leads to decays suppressed relative to main ones, as
it has been explained in the previous section. Therefore, we concentrate
here on weak decays $T_{b:\overline{d}}^{-}\rightarrow \widetilde{Z}%
_{bc}^{0}M$ of the tetraquark $T_{b:\overline{d}}^{-}$.

In these processes $M$ is one of the pseudoscalar mesons $\pi ^{-}$, $K^{-}$%
, $\ D^{-}$, and $\ D_{s}^{-}$. They appear at the final state due to decays
of $W^{-}$ to quark-antiquark pairs $d\overline{u}$, $s\overline{u}$, $d%
\overline{c}$, and $s\overline{c}$, respectively. In Table \ref{tab:MesonPar}
we present the masses and decay constants of the mesons $\pi ^{-}$, $K^{-}$,
$\ D^{-}$, and $\ D_{s}^{-}$. It is easy to see, that the mass of the master
particle $T_{b:\overline{d}}^{-}$ meets a requirement $m>\widetilde{m}%
_{Z}+m_{M}$, and all these decays are kinematically allowed processes.

It is convenient to describe production of mesons $M$ using the effective
Hamiltonian, and introduce relevant effective weak vertices. We restrict
ourselves by analyzing only tree-level contributions to decays: the relevant
Feynman diagram for the process $T_{b:\overline{d}}^{-}\rightarrow
\widetilde{Z}_{bc}^{0}K^{-}$, as an example, is depicted in Fig.\ \ref%
{fig:NLDecay}. To study the nonleptonic weak decays $T_{b:\overline{d}%
}^{-}\rightarrow \widetilde{Z}_{bc}^{0}M$, we also adopt the QCD
factorization method. This approach was applied to investigate nonleptonic
decays of conventional mesons \cite{Beneke:1999br,Beneke:2000ry}, but can be
used to study decays of the tetraquarks as well. Thus, nonleptonic decays of
the scalar exotic mesons $Z_{bc}^{0}$, $T_{bs;\overline{u}\overline{d}}^{-}$
and $T_{bb;\overline{u}\overline{s}}^{-}$ (in a short form $T_{b:\overline{s}%
}^{-}$) were analyzed by this way in Refs.\ \cite%
{Sundu:2019feu,Agaev:2019wkk,Agaev:2019lwh}, respectively.

We provide details of analysis for the decay $T_{b:\overline{d}%
}^{-}\rightarrow \widetilde{Z}_{bc}^{0}\pi ^{-}$, and write down final
predictions for other channels.

\begin{figure}[h]
\includegraphics[width=8.8cm]{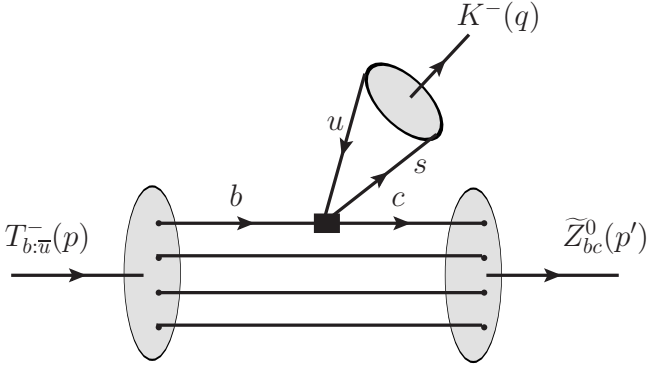}
\caption{The tree-level Feynman diagram for the nonleptonic decay $T_{b:%
\overline{d}}^{-}\rightarrow \widetilde{Z}_{bc}^{0}K^{-}$. The black square
denotes the effective weak vertex.}
\label{fig:NLDecay}
\end{figure}
At the tree-level, the effective Hamiltonian for this decay is given by the
expression
\begin{equation}
\mathcal{H}_{\mathrm{n.-lep}}^{\mathrm{eff}}=\frac{G_{F}}{\sqrt{2}}%
V_{bc}V_{ud}^{\ast }\left[ c_{1}(\mu )Q_{1}+c_{2}(\mu )Q_{2}\right] ,
\label{eq:EffHam}
\end{equation}%
where%
\begin{eqnarray}
Q_{1} &=&\left( \overline{d}_{i}u_{i}\right) _{\mathrm{V-A}}\left( \overline{%
c}_{j}b_{j}\right) _{\mathrm{V-A}},  \notag \\
Q_{2} &=&\left( \overline{d}_{i}u_{j}\right) _{\mathrm{V-A}}\left( \overline{%
c}_{j}b_{i}\right) _{\mathrm{V-A}},  \label{eq:Operators}
\end{eqnarray}%
and $i$ , $j$ are the color indices, and $\left( \overline{q}%
_{1}q_{2}\right) _{\mathrm{V-A}}$ means
\begin{equation}
\left( \overline{q}_{1}q_{2}\right) _{\mathrm{V-A}}=\overline{q}_{1}\gamma
_{\mu }(1-\gamma _{5})q_{2}.  \label{eq:Not}
\end{equation}%
It is worth noting that, we do not include into Eq.\ (\ref{eq:EffHam})
current-current operators appearing due the QCD penguin and
electroweak-penguin diagrams. The short-distance Wilson coefficients $%
c_{1}(\mu )$ and $c_{2}(\mu )$ are given at the factorization scale $\mu $.

In the factorization method the amplitude of the decay $T_{b:\overline{d}%
}^{-}\rightarrow \widetilde{Z}_{bc}^{0}\pi ^{-}$ has the form
\begin{eqnarray}
\mathcal{A} &=&\frac{G_{F}}{\sqrt{2}}V_{bc}V_{ud}^{\ast }a_{1}(\mu )\langle
\pi ^{-}(q)|\left( \overline{d}_{i}u_{i}\right) _{\mathrm{V-A}}|0\rangle
\notag \\
&&\times \langle \widetilde{Z}_{bc}^{0}(p^{\prime })|\left( \overline{c}%
_{j}b_{j}\right) _{\mathrm{V-A}}|T_{b:\overline{d}}^{-}(p)\rangle,
\label{eq:Amplitude}
\end{eqnarray}%
where
\begin{equation}
a_{1}(\mu )=c_{1}(\mu )+\frac{1}{N_{c}}c_{2}(\mu ),
\end{equation}%
with $N_{c}=3$ being the number of quark colors. The only unknown matrix
element $\langle \pi ^{-}(q)|\left( \overline{d}_{i}u_{i}\right) _{\mathrm{%
V-A}}|0\rangle $ in $\mathcal{A}$ can be defined in the following form

\begin{equation}
\langle \pi ^{-}(q)|\left( \overline{d}_{i}u_{i}\right) _{\mathrm{V-A}%
}|0\rangle =if_{\pi }q_{\mu }.  \label{eq:ME4}
\end{equation}%
Then, it is not difficult to see that $\mathcal{A}$ is%
\begin{eqnarray}
\mathcal{A} &=&i\frac{G_{F}}{\sqrt{2}}f_{\pi }V_{bc}V_{ud}^{\ast }a_{1}(\mu )
\notag \\
&&\times \left[ G_{1}(q^{2})Pq+G_{2}(q^{2})q^{2}\right].
\label{eq:Amplitude2}
\end{eqnarray}%
For completeness we provide below the partial width of this process%
\begin{eqnarray}
&&\Gamma (T_{b:\overline{d}}^{-}\rightarrow \widetilde{Z}_{bc}^{0}\pi ^{-})=%
\frac{G_{F}^{2}f_{\pi }^{2}|V_{bc}|^{2}|V_{ud}|^{2}}{32\pi m^{3}}  \notag \\
&&\lambda \left( m^{2},\widetilde{m}_{Z}^{2},m_{\pi }^{2}\right) \left[
G_{1}(m^{2}-\widetilde{m}_{Z}^{2})+G_{2}m_{\pi }^{2}\right] ^{2},
\label{eq:NLDW}
\end{eqnarray}%
where the weak form factors $G_{1(2)}(q^{2})$ are computed at $q^{2}=m_{\pi
}^{2}$. The decay modes $T_{b:\overline{d}}^{-}\rightarrow \widetilde{Z}%
_{bc}^{0}K^{-}(D^{-},\ D_{s}^{-})$ can be analyzed in a similar manner. To
this end, one has to replace in Eq.\ (\ref{eq:NLDW}) ($m_{\pi },f_{\pi }$)
by the masses and decay constants of the mesons $K^{-}$, $D^{-}$, and $%
D_{s}^{-}$,  make the substitutions $V_{ud}\rightarrow V_{us}$, $V_{cd}$,
and $V_{cs}$,  and fix the form factors at $q^2=m_{M}^{2}$.

All input information necessary for numerical analysis are collected in
Table \ref{tab:MesonPar}: it contains spectroscopic parameters of the final
state mesons, and CKM matrix elements. The coefficients $c_{1}(m_{b})$, and $%
c_{2}(m_{b})$ with next-to-leading order QCD corrections are borrowed from
Refs.\ \cite{Buras:1992zv,Ciuchini:1993vr,Buchalla:1995vs}
\begin{equation}
c_{1}(m_{b})=1.117,\ c_{2}(m_{b})=-0.257.  \label{eq:WCoeff}
\end{equation}

\begin{table}[tbp]
\begin{tabular}{|c|c|}
\hline\hline
Quantity & Value \\ \hline\hline
$m_{\pi} $ & $139.570~\mathrm{MeV}$ \\
$m_{K}$ & $(493.677\pm 0.016)~\mathrm{MeV}$ \\
$m_{D}$ & $(1869.61 \pm 0.10)~\mathrm{MeV}$ \\
$m_{D_s}$ & $(1968.30\pm 0.11)~\mathrm{MeV}$ \\
$f_{\pi }$ & $131~\mathrm{MeV}$ \\
$f_{K}$ & $(155.72\pm 0.51)~\mathrm{MeV}$ \\
$f_{D}$ & $(203.7 \pm 4.7)~\mathrm{MeV}$ \\
$f_{D_s}$ & $(257.8 \pm 4.1)~\mathrm{MeV}$ \\
$|V_{ud}|$ & $0.97420\pm 0.00021$ \\
$|V_{us}|$ & $0.2243\pm 0.0005$ \\
$|V_{cd}|$ & $0.218\pm 0.004$ \\
$|V_{cs}|$ & $0.997\pm 0.017$ \\ \hline\hline
\end{tabular}%
\caption{Masses and decay constants of the final state pseudoscalar mesons.
The CKM matrix elements are also included. }
\label{tab:MesonPar}
\end{table}

For the decay $T_{b:\overline{d}}^{-}\rightarrow \widetilde{Z}_{bc}^{0}\pi
^{-}$, calculations yield
\begin{eqnarray}
\Gamma (T_{b:\overline{d}}^{-} &\rightarrow &\widetilde{Z}_{bc}^{0}\pi
^{-})=\left( 5.13\pm 1.42\right) \times 10^{-13}~\mathrm{MeV}.  \notag \\
&&  \label{eq:NLDW1}
\end{eqnarray}%
Partial widths of this and other nonleptonic decays of the tetraquark $T_{b:%
\overline{d}}^{-}$ are moved to Table \ref{tab:PWidth}. It is evident that
widths of these processes are very small, and can be safely neglected in
computation of the full width of the $T_{b:\overline{d}}^{-}$.

As a result, we get
\begin{eqnarray}
\Gamma _{\mathrm{full}} &=&(10.88\pm 1.88)\times 10^{-10}~\mathrm{MeV},
\notag \\
\tau &=&6.05_{-0.89}^{+1.26}\times 10^{-13}~\mathrm{s},  \label{eq:WL}
\end{eqnarray}%
which are among main predictions of the present work.

%%%%%%%%%%%%%%%%%%%%%%%%%%%%%%%%%%%%%%%%%%%%%%%%%%%%%%%%%%%%%%%%%%%%%%%%%%%%

\section{Analysis and concluding remarks}

\label{sec:Disc}
%%%%%%%%%%%%%%%%%%%%%%%%%%%%%%%%%%%%%%%%%%%%%%%%%%%%%%%%%%%

In the present work we have calculated the mass, width and lifetime of the
stable scalar tetraquark $T_{b:\overline{d}}^{-}$ with the content $bb%
\overline{u}\overline{d}$. This particle can be considered as a $ud$ member
of the scalar multiplet $bb\overline{q}\overline{q}^{\prime }$. Another
particle from this multiplet $T_{b:\overline{s}}^{-}$ was studied in our
article \cite{Agaev:2019lwh}. The tetraquark $T_{b:\overline{s}}^{-}$ is
composed of $bb\overline{u}\overline{s}$ quarks, has the mass
\begin{equation}
m=(10250~\pm 270)~\mathrm{MeV,}
\end{equation}%
and is stable against the strong and electromagnetic decays. By comparing
parameters of the tetraquarks $T_{b:\overline{s}}^{-}$ and $T_{b:\overline{d}%
}^{-}$ one can easily reveal a mass gap $115~\mathrm{MeV}$ in this
multiplet, which is consistent with analysis of the open charm-bottom
axial-vector states $Z_{s}=[cs][\overline{b}\overline{s}]$ and $Z_{q}=[cq][%
\overline{b}\overline{q}]$ \cite{Agaev:2017uky}. In fact, the mass splitting
between $Z_{s}$ and $Z_{q}$ equals approximately to $240~\mathrm{MeV}$,
which is caused by two $s$ quarks in the $Z_{s}$, hence a single $s$
generates the mass splitting $120~\mathrm{MeV}$.

The second particle considered in this work is the tetraquark $\widetilde{Z}%
_{bc}^{0}$ appeared due to weak decays of the master particle $T_{b:%
\overline{d}}^{-}$. We have treated $\widetilde{Z}_{bc}^{0}$ as a scalar
exotic meson $bc\overline{u}\overline{d}$ built of diquark and antidiquark
with symmetric color structures, and calculated its spectroscopic parameters
$\widetilde{m}_{Z}$ and $\widetilde{f}_{Z}$. The scalar particle $\widetilde{%
Z}_{bc}^{0}$ is stable against $S$-wave decays to mesons $B^{-}D^{+}$ and $%
\overline{B}^{0}D^{0}$ because thresholds for these processes $7149/7144~%
\mathrm{MeV}$ are higher than mass of the $\widetilde{Z}_{bc}^{0}$. For the
same reasons $\widetilde{Z}_{bc}^{0}$ can not transform to conventional
mesons through electromagnetic decays. In fact, threshold for a such process
$\widetilde{Z}_{bc}^{0}\rightarrow B^{-}D_{s1}(2460)^{+}\gamma $ is equal to
$7739~\mathrm{MeV}$ and considerably exceeds the mass of the tetraquark $%
\widetilde{Z}_{bc}^{0}$.

There are two other scalar exotic mesons with the same or close quark
contents. First of them is particle $Z_{bc}^{0}=bc\overline{u}\overline{d}$
composed of the color-triplet diquark and antidiquark. The mass of this
exotic meson is equal to $m_{Z}=(6660\pm 150)~\mathrm{MeV}$ \cite%
{Agaev:2018khe}. The second scalar tetraquark is $s$ partner of $\widetilde{Z%
}_{bc}^{0}$, i.e., an exotic meson $Z_{b:\overline{s}}^{0}=bc\overline{u}%
\overline{s}$ with color-sextet organization of constituent diquarks. This
particle was investigated in Ref.\ \cite{Agaev:2019lwh}, in which its mass
was estimated within the range
\begin{equation}
\widetilde{m}=(6830~\pm 140)~\mathrm{MeV.}
\end{equation}

The mass splitting inside of the multiplet of scalar particles $bc\overline{q%
}\overline{q}^{\prime }$ with color-sextet structure of diquark and
antidiquark
\begin{equation}
\widetilde{m}-\widetilde{m}_{Z}=100~\mathrm{MeV,}  \label{eq:Split2}
\end{equation}%
is compatible with our above-stated discussion. Comparing the masses of $%
\widetilde{Z}_{bc}^{0}$ and $Z_{bc}^{0}$ with the color-sextet and -triplet
organization of constituents, we get
\begin{equation}
\Delta m=\widetilde{m}_{Z}-m_{Z}=70~\mathrm{MeV.}  \label{eq:Split1}
\end{equation}%
The mass gap between axial-vector four-quark mesons $[cs][\overline{c}%
\overline{s}]$ with different color structures of constituent diquarks was
studied in Ref.\ \cite{Agaev:2017foq}. The "color-triplet" and "color-sextet"
states were interpreted there as candidates to resonances $X(4140)$ and $%
X(4274)$, respectively. The theoretical estimate for a difference of their
masses amounts to $\Delta m\approx 180~\mathrm{MeV}$. The triplet-sextet
splitting in the scalar system $(Z_{bc}^{0},\widetilde{Z}_{bc}^{0})$ is
numerically smaller than in the case of axial-vector tetraquarks. But one
should take into account that axial-vector particles $[cs][\overline{c}%
\overline{s}]$ are composed of a heavy diquark and an antidiquark, whereas
tetraquarks $bc\overline{q}\overline{q}^{\prime }$ are built of the heavy
diquark and light antidiquark. Whether the triplet-sextet splitting depends
only on spin-parities of these particles or bears also information on their
structures, worths additional studies.

The estimates presented above for splitting of different tetraquarks are
found using central values of their masses. Parameters of these states,
including their masses, have been extracted by means of the QCD sum rule
method, predictions of which contain theoretical uncertainties. Therefore,
the results for mass splitting in the multiplet of double-heavy tetraquarks
should be considered with some caution. In our view, the picture drawn
above, nevertheless, is a credible image of the real exotic-meson
spectroscopy.

We have computed partial widths of the semileptonic $T_{b:\overline{d}%
}^{-}\rightarrow \widetilde{Z}_{bc}^{0}l\overline{\nu }_{l}$ and nonleptonic
$T_{b:\overline{d}}^{-}\rightarrow \widetilde{Z}_{bc}^{0}M$ decays, where $M$
is one of the pseudoscalar mesons $\pi ^{-}$, $K^{-}$, $D^{-}$, and $%
D_{s}^{-}$. In these processes final hadronic states are either the scalar
tetraquark $\widetilde{Z}_{bc}^{0}$ or this tetraquark and a conventional
meson $M$. It turned out that partial widths of semileptonic decays are
considerably higher than ones of nonleptonic modes. Namely the semileptonic
decay channels have been used to evaluate the full width and lifetime of $%
T_{b:\overline{d}}^{-}$. It should be noted that there are weak nonleptonic
decays of $T_{b:\overline{d}}^{-}$ which at the final state contains two
ordinary mesons. Such processes were analyzed in Ref.\ \cite{Ali:2018ifm},
in which the authors considered decays of the axial-vector tetraquark $%
T_{bb}^{-}$. Similar channels can be examined in the case of the scalar
particle $T_{b:\overline{d}}^{-}$ as well. But, partial widths of these
modes are considerably smaller than widths of the semileptonic decays, and
latter determine mean lifetime of $T_{b:\overline{d}}^{-}$.

Till now the experimental collaborations did not observe weakly decaying
tetraquarks, which would be strong evidence for their existence. It is worth
noting that active experiments, such as LHCb, have a certain potential to
discover weak decay modes of tetraquarks $T_{bb}$. Such potential should
have also a Tera-$Z$ factory. In Refs.\ \cite{Ali:2018ifm,Ali:2018xfq} the
authors addressed namely these problems, and considered the processes
$Z\rightarrow b\overline{b}b\overline{b}$,
$pp\rightarrow b\overline{b}b\overline{b}+X$, and $pp\rightarrow b\overline{b}c%
\overline{c}+X$ to estimate
production rates of double heavy tetraquarks. It was found that the
integrated cross section for production of $T_{bb}^{-}$ is
\begin{equation}
\sigma \left( pp\rightarrow T_{bb}^{-}+X\right) =2.8_{-0.7}^{+1.0}~\mathrm{nb%
},
\end{equation}%
whereas for the tetraquark with the content $T_{bc}^{0}=bc\overline{u}%
\overline{d}$ similar analysis leads to estimate
\begin{equation}
\sigma \left( pp\rightarrow T_{bc}^{0}+X\right) =103_{-25}^{+39}~\mathrm{nb}.
\end{equation}%
In accordance with predictions of Ref.\ \cite{Ali:2018xfq}, this implies
producing of approximately $\mathcal{O}(10^{8})$ events with $T_{bb}^{-}$
and $\mathcal{O}(10^{9})$ events with $T_{bc}^{0}$ during LHC Runs $1-4$.

The $Z$-boson factories with the integrated luminosity of $10^{12}$ $Z$%
-boson events may lead to production significant number of tetraquarks $%
T_{bb}^{-}$ and allow one to measure its parameters. This conclusion is
based on the estimate for the branching ratio%
\begin{equation}
\mathcal{B}(Z\rightarrow T_{bb}^{-}+\overline{b}\overline{b})=\left(
1.4_{-0.5}^{+1.1}\right) \times 10^{-6}
\end{equation}%
made in Ref.\ \cite{Ali:2018ifm}.

The production of the tetraquarks $T_{b:\overline{d}}^{-}$, $\widetilde{Z}%
_{bc}^{0}$, and $Z_{bc}^{0}$ in proton-proton collisions at LHC and future $%
Z $-factories seems may be analyzed within the scheme discussed in Refs.\
\cite{Ali:2018xfq,Ali:2018ifm} by taking into account differences due to
scalar nature of these particles. One also can utilize the QCD sum rule
method to evaluate some of matrix elements used in these investigations and
refine existing approach. Relevant processes in $pp$ and $e^{+}e^{-}$%
collisions require detailed studies and analysis, which are beyond the scope
of the present work.

Spectroscopic parameters of the scalar particles $T_{b:\overline{d}}^{-}$
and $\widetilde{Z}_{bc}^{0}$, as well as weak decays of $T_{b:\overline{d}%
}^{-}$ studied in the present work provide new and useful information on
features of double-heavy exotic mesons $bb\overline{q}\overline{q}^{\prime }$
and $bc\overline{q}\overline{q}^{\prime }$, and form a basis for future
investigations.

\section*{ACKNOWLEDGEMENTS}

The work of K.~A, B.~B., and H.~S was supported in part by the TUBITAK grant
under No: 119F050.

\end{document}